\documentclass[12pt]{article} 

\usepackage{graphicx}
\usepackage{array,mathptm,amsmath,vmargin,tabularx}
\usepackage{dcolumn,booktabs}
\usepackage{multicol}
\newcolumntype{d}[1]{D{.}{.}{#1}}

\begin{document}

\title{Percolation in suspensions of hard nanoparticles: From spheres to needles}

\author{Tanja Schilling\\ Research Unit for Physics and Materials Science\\ Universit\'e du Luxembourg\\ L-1511 Luxembourg, Luxembourg\\ \and Mark A. Miller\\ Department of Chemistry, Durham University\\ South Road, Durham DH1 3LE, United Kingdom\\ \and Paul van der Schoot\\Theory of Polymers and Soft Matter\\ Technische Universiteit Eindhoven\\ Postbus 513, 5600 MB Eindhoven, The Netherlands}

\maketitle

\abstract{We investigate geometric percolation and scaling relations in suspensions of nanorods, covering the entire range of aspect ratios from spheres to extremely slender needles.  A new version of connectedness percolation theory is introduced and tested against specialised Monte Carlo simulations.  The theory accurately predicts percolation thresholds for aspect ratios of rod length to width as low as 10.  The percolation threshold for rod-like particles of aspect ratios below 1000 deviates significantly from the inverse aspect ratio scaling prediction, thought to be valid in the limit of infinitely slender rods and often used as a rule of thumb for nano-fibres in composite materials. Hence, most fibres that are currently used as fillers in composite materials cannot be regarded as practically infinitely slender for the purposes of percolation theory. Comparing percolation thresholds of hard rods and new benchmark results for ideal rods, we find that (i) for large aspect ratios, they differ by a factor that is inversely proportional to the connectivity distance between the hard cores, and (ii) they approach the slender rod limit differently.}

\newpage
Connectivity percolation is the transition in which isolated
clusters of solid particles in a fluid (or of voids in a solid) become connected in some sense to form a system-spanning network. This network has a significant effect on the mechanical and transport properties of the material on a macroscopic scale \cite{torquato:2002}. If, for example, an electrically insulating polymer is mixed with conductive fibres such as carbon nanotubes, the conductivity of the composite increases by ten or more orders of magnitude near the percolation transition of the filler material \cite{nan:2010}. Given the technological relevance to opto-electronics, photovoltaics and electromagnetic radiation shielding, it is no surprise that a large research effort is currently being invested to understand how the formulation and processing of a composite influence the percolation threshold as well as its physical properties beyond the threshold \cite{park:2013}.

The topic of percolation originates from studies on
fluid flow in porous media, relevant, for example, to oil extraction. It has since been extensively studied theoretically and computationally, both on and off lattice, the latter particularly (but not exclusively) for ideal, non-interacting bodies. An important scientific motivation for these studies is the critical behavior that the percolation transition shares with phase transitions \cite{aharony:2003}. For spherical particles, the impact of repulsive and attractive interactions on continuum percolation has received considerable attention \cite{torquato:2002}, while for non-spherical particles such as nano-wires, current understanding is much sketchier, despite their industrial interest as fillers in composite materials.

Fibre-like fillers have been modelled as
cylinders, spherocylinders and ellipsoids in theoretical studies \cite{kyrylyuk:2008, yi:2004, wang:2003, leung:1991, chatterjee:2000, munson-mcgee:1991, celzard:1996}, in simulations where interactions are ignored \cite{pike:1974, neda:1999, foygel:2005, mutiso:2012}, and in simulations where the particles interact via a hard excluded volume \cite{berhan:2007, schilling:2007, schilling:2009, ambrosetti:2010}.
However, there is no systematic test of theory against simulation over a large range of aspect ratios for interacting particles. In particular, the intermediate regime, in which the length is tens or a few hundred times the thickness, has not been addressed yet, even though most fibres used in realistic materials fall into this range. The reasons for this are that it is very time-consuming to simulate interacting particles of high length-to-width aspect ratios, and that analytical theories that are thought to be accurate in the limit of infinite aspect ratio are difficult to extend to finite values \cite{otten:2011}.

In this letter we present a combined theoretical and simulation study of percolation in suspensions of hard spherocylinders that spans an unprecedented range of aspect ratios from spheres to slender rods of aspect ratio up to 1000.  We show that hard-core interactions (i) change the approach to the theoretically expected scaling of the percolation threshold with the inverse aspect ratio, (ii) shift the percolation threshold to larger values than that for ideal rods by a factor that converges only slowly with respect to rod length and (iii) cause the core packing fraction at percolation to exhibit a maximum for small aspect ratios. These results show that the ideal (penetrable) particle model has limited predictive value for actual rod systems and that even for very long rods, a finite-length correction is necessary to get quantitative predictions. We show that this correction can be obtained explicitly from connectedness percolation theory, using a sensible ansatz for the connectedness direct correlation function.

\begin{figure}
\begin{center}
  \includegraphics[width=0.5\columnwidth]{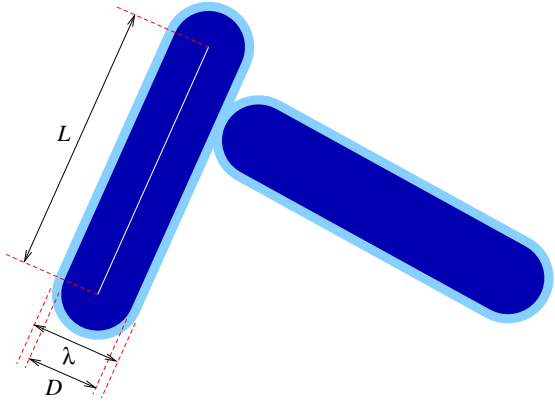}
\end{center}
  \caption{Schematic illustration of two spherocylinders with overlapping contact shells.  $L$ is the length of the line segment (white) of each spherocylinder, which is common to the hard core and to the contact shell, $D$ is the diameter of the hard core (dark), and $\lambda$ is the diameter of the contact shell (light).
    \label{fig:scdef}}
\end{figure}

Before discussing the results, we briefly describe our simulation method and theory.  We have generated configurations of hard spherocylinders at fixed particle number $N$ and volume $V$, using cubic simulation cells of length $L_x=V^{1/3}$. The spherocylinders consist of a cylinder of length $L$ and diameter $D$, capped with hemispheres of the same diameter.  Hence, the surface of the spherocylinder consists of all points lying a distance $D/2$ from a line segment of length $L$.  The full aspect ratio of a hard spherocylinder, including the caps, is $L/D+1$.
The core is strictly impenetrable but is surrounded by a notional contact shell that is used to define when two spherocylinders are considered to be connected.  The surface of the contact shell is a spherocylinder that shares the same line segment as the core but has diameter $\lambda$ instead of $D$.  Hence, the surface of the contact shell lies at a uniform distance $(\lambda - D)/2$ from the surface of the core (see Fig.~\ref{fig:scdef}).  The full aspect ratio of the contact shell is therefore $L/\lambda+1$.  Two hard spherocylinders are connected if their shells overlap, and clusters are defined by contiguous pairwise connections.

\begin{figure}
  \includegraphics[width=0.9\columnwidth]{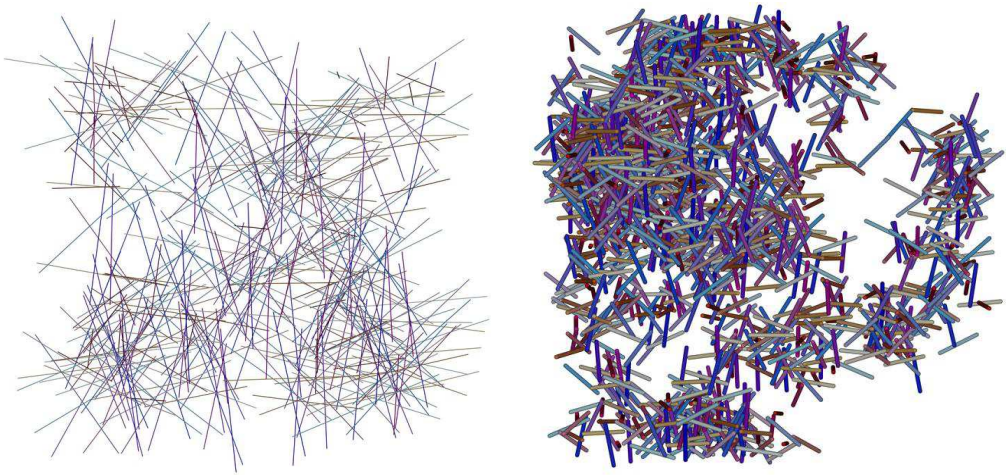}
  \caption{Simulation snapshots of percolating clusters. The rods are colour coded according to their orientation. Left panel: $L/D = 200$, right panel: $L/D=10$.
    \label{fig:snapshots}}
\end{figure}

We sample the fraction $P(\phi;L_x)$ of independent configurations that contain a percolating cluster as a function of the packing fraction $\phi$ of the cores for a given cell length $L_x$.  The packing fraction is defined by $\phi = Nv_{\rm core}/V$, where $v_{\rm core} = \pi D^2\left(\frac{L}{4} + \frac{D}{6}\right)$ is the volume of one spherocylinder.
To detect a percolating cluster in a cubic simulation cell with periodic boundary conditions, we require that a cluster must connect periodic images of its constituent particles in at least one of the periodic directions. This ``wrapping'' criterion is somewhat more costly to evaluate than the simpler ``spanning'' criterion, which only requires that a cluster connects two opposite boundaries of the simulation cell.  However, wrapping clusters are a more accurate representation of a percolating cluster in the macroscopic limit because such clusters are infinite when the simulation cell is replicated through its boundary conditions.  In contrast, spanning clusters merely form an array of large but disconnected clusters when the cell is replicated.  Furthermore, wrapping probabilities follow universal scaling functions \cite{Skvor07a}---a feature that we will exploit in the simulations of ideal rods to mitigate the small systematic errors arising from the inevitably finite size of the simulation cell.  At finite $L_x$,
$P(\phi;L_x)$ is a sigmoidal function of $\phi$, becoming a sharp step function as $L_x\to\infty$. The curves for different $L_x$ have a common crossing point \cite{Skvor07a}, typically just below $P=0.5$.
To make the simulations tractable, we identify the hard-rod percolation threshold $\phi_{\rm p}$ as the point $P(\phi_{\rm p};L_x)=0.5$ using a cell length up to $L_x=10L$ (never less than $L_x=15D$). Fig.~\ref{fig:snapshots} shows examples of percolating clusters for $L/D=200$ and $L/D=10$.

Properly equilibrated configurations of impenetrable rods cannot be generated by sequential random insertion \cite{Miller09c}. Configurations must therefore be obtained by Monte Carlo displacements and rotations of particles, rejecting any trial move that would generate an overlap of cores and accepting all others. This procedure becomes computationally costly for slender rods, where the simulation cell and number of particles must be large to allow for an accurate computation of the percolation threshold.
To find overlaps quickly, we use a method in which the spherocylinders are notionally divided into small segments so that the overlap detection need only be performed for neighboring segments \cite{vink:2005}.  This method scales linearly with particle number but requires a large amount of memory.

We have also calculated the percolation threshold for fully penetrable (ideal) spherocylinders, consisting of the contact shell with no hard core.  By definition, $\phi=0$ for ideal rods, but hard and ideal spherocylinders may be compared using a packing fraction renormalised to the volume of the contact shell, $\eta=Nv_{\rm shell}/V$, where $v_{\rm shell} = \pi\lambda^2\left(\frac{L}{4} + \frac{\lambda}{6}\right)$.  $\eta$ is the hypothetical volume fraction of the shells in the absence of shell overlaps.  Independent equilibrium configurations of ideal spherocylinders are easily generated by placing rods with random positions and orientations within the simulation cell.  For ideal rods we used cells lengths up to $L_x=6L$ (never less than $16\lambda$) and identified the percolation threshold from the crossing point of $P(\eta;L_x)$ at two cell sizes $L_x$, sampling 20\,000 configurations at each $\eta$ and $L_x$.  This procedure minimises the systematic errors due to finite cell sizes and reduces statistical uncertainty in the ideal-rod percolation thresholds to $0.1\%$.

Our theoretical predictions are based on connectedness percolation theory.  The percolation threshold is defined as the filler fraction $\phi$ at which the mean cluster size diverges and is equal to $v_{\rm core}/\langle\langle \hat{C}^+ \rangle \rangle'$ \cite{otten:2011}. Here, $v_{\rm core}$ is again the particle volume, $\hat{C}^+$ the spatial Fourier transform of the connectedness direct correlation function, $C^+ (\mathbf{r},\mathbf{u},\mathbf{u}')$, at zero wave vector. The direct correlation function is a function of the vector $\mathbf{r}$ connecting the centres of mass of the particles, and their main-body-axis vectors $\mathbf{u}$ and $\mathbf{u}'$. The angular brackets denote an orientational averaging over these two directors. In the limit of long thin rods the second virial approximation is accurate \cite{otten:2011}. Within the second virial approximation, $C^+ (\mathbf{r},\mathbf{u},\mathbf{u}')=f^+ (\mathbf{r},\mathbf{u},\mathbf{u}')$, where $f^+ \equiv \exp (- U^+)$ is the connectedness Mayer function with $U^+=U^+(\mathbf{r},\mathbf{u},\mathbf{u}')$ the connectedness interaction potential scaled to the thermal energy. $U^+=\infty$ for non-connected configurations and $U^+=0$ for connecting ones, i.e., configurations for which the connectivity shells overlap but the hard cores do not.

To go beyond the second virial approximation, we invoke a Lee--Parsons type of approximation that has proven remarkably accurate in predicting the phase behavior of hard rods and mixtures of hard rods and hard spheres \cite{cuetos:2007}. It is based on an interpolation between the Percus--Yevick equation of state for hard spheres and the second virial equation of state for hard rods. In the context of connectedness percolation, it can be written as $C^+=f^+ \times (1-3\phi/4)/(1-\phi)^2$, given the known relations between the direct correlation function and the connectedness variant of it \cite{otten:2011, vanderschoot:1996}. This then gives for the percolation threshold an explicit expression in terms of the ratio $\gamma = L/D$ and the dimensionless measure of the connectivity range $\alpha = \lambda/D - 1$,
\begin{equation}
\phi_{\rm p} = \frac{2\left(1+\xi\right) - 2 \left(1+\frac{1}{2}\xi\right)^\frac{1}{2}}{3\left(1+\frac{2}{3}\xi\right)}, \label{eq:eta}
\end{equation}
where
\begin{equation}
\xi(\gamma, \alpha) = \frac{\left(1+\frac{2}{3}\frac{1}{\gamma}\right)\frac{1}{\gamma}}{\frac{8}{3\gamma^2}\left(\left(1+\alpha\right)^3-1\right)+\frac{4}{\gamma}\left(\left(1+\alpha\right)^2-1\right)+\alpha}.
\end{equation}

To obtain numerical results for the percolation threshold, we must select a value for the connectivity criterion $\lambda$.  The separation $\lambda-D$ of the core spherocylinder surfaces at the connectivity cut-off (see Fig.~\ref{fig:scdef}) should be characteristic of the distance over which electron tunnelling between nanorods decays \cite{Hu06}.  This distance depends on the details of both the nanorods and the medium in which they are suspended \cite{kyrylyuk:2008} and a full quantum mechanical treatment is a formidable task.  However, if the energetic barrier $\Delta E$ for tunnelling can be measured or calculated then an estimate for $\lambda-D$ can be obtained from the tunnelling length $(\hbar^2/2m_{\rm e}\Delta E)^{1/2}$ through a rectangular barrier, where $m_{\rm e}$ is the electron mass.  For suspensions of carbon nanotubes (diameters on the order of a nanometre), the tunnelling length typically lies at a fraction of a nanometre \cite{kyrylyuk:2008}.  We have therefore taken $\lambda/D=1.2$ as a representative value for most of our results, but will consider the effects of altering this value later.

\begin{figure}
  \includegraphics[width=0.9\columnwidth]{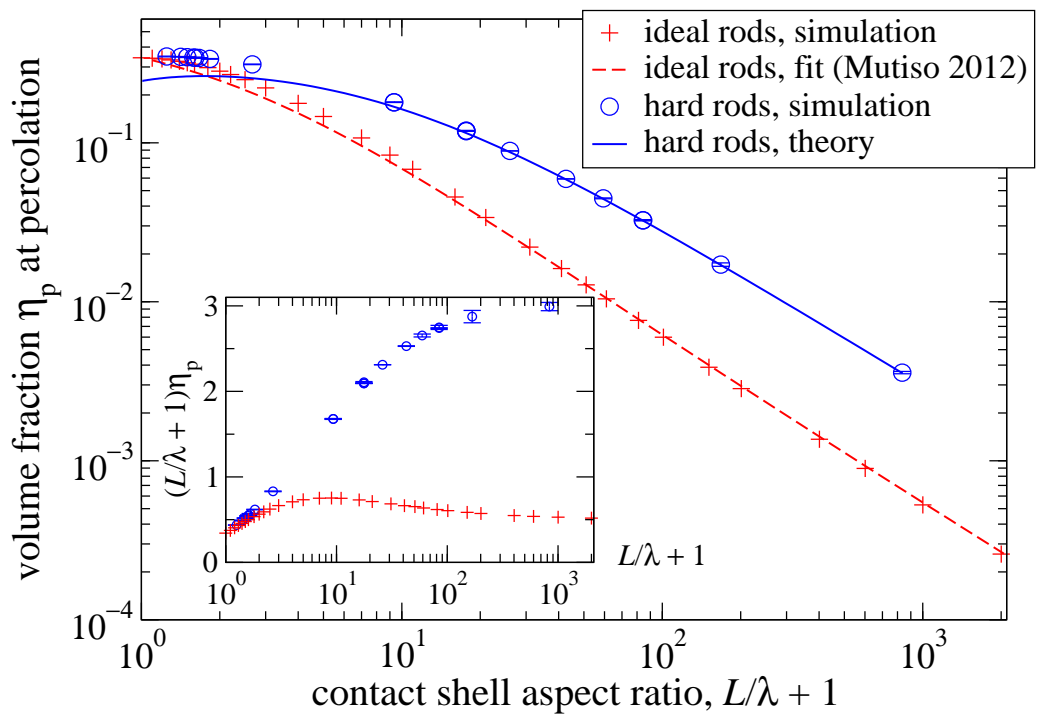}
  \caption{Percolation threshold $\eta_{\rm p}$ as a function of aspect ratio for ideal rods and hard rods.  The approach to the scaling regime, where $(L/\lambda+1)\eta$ would be constant in each case, is highlighted in the inset.  Statistical errors in the ideal-rod results are smaller than the symbols. Tabulated data for the main plot are available as supplementary material.
    \label{fig:Allthresholds}}
\end{figure}

In Fig.~\ref{fig:Allthresholds} we present our simulation results (circles) for the percolation threshold of hard rods in terms of the renormalised volume fraction $\eta$ versus the full aspect ratio $L/\lambda+1$, from spheres (aspect ratio 1, $L/D=0$) to very slender rods with core $L/D=1000$.  For comparison, we also show (i) our simulation results for ideal rods (pluses), (ii) the phenomenological expression for ideal cylinders of the same aspect ratio, obtained by Mutiso et al.~using a fit to simulation data on somewhat shorter rods \cite{mutiso:2012} (dashed line), and (iii) our theoretical prediction (solid line). Our theory agrees quantitatively with the simulations of hard rods for aspect ratios above 10, and semi-quantitatively below that.
The largest discrepancy between simulation and theory occurs for aspect ratios approaching zero and amounts to $25$ \% for our choice of connectivity range. This is to be expected given the level of accuracy of the Percus--Yevick approximation for the percolation of hard spheres \cite{desimone:1986}. The Mutiso fit for ideal cylinders\cite{mutiso:2012} captures the qualitative deviations from the long-rod limit, but only agrees with our simulations within statistical error over a narrow middle range of aspect ratios.

A number of conclusions can be drawn from Fig.~\ref{fig:Allthresholds}.  First, we may compare the accurate results of the Monte Carlo simulations with the asymptotic scalings for the percolation thresholds of long rods.  The long-rod scaling predictions are $\eta_{\rm p} \sim \lambda/2L$ for ideal rods \cite{Bug85b} and $\eta_{\rm p} \sim \lambda^2/2L(\lambda-D)$ for hard rods.  Both these results can be obtained from connectedness percolation theory within the second virial approximation in the limit $L\gg D$ \cite{kyrylyuk:2008}.  However, the same prediction for ideal rods was also made in earlier work \cite{Bug85b,Bug86a}, using a conjecture based directly on average excluded volume and the number of contacts between objects at the percolation threshold.
In both the ideal- and hard-rod cases, the product $(L/\lambda+1)\eta$ should approach a constant value with increasing rod length.  In the inset of Fig.~\ref{fig:Allthresholds}, however,
we see that the asymptotic scalings for the percolation threshold of long rods are only reached for aspect ratios in excess of several hundred \cite{otten:2011}.
For hard rods, a constant value of $(L/\lambda+1)\eta$ is reached slowly from below, while for ideal rods the respective plateau is approached from above after initially overshooting.  

\begin{figure}
  \centering
  \includegraphics[width=0.9\columnwidth]{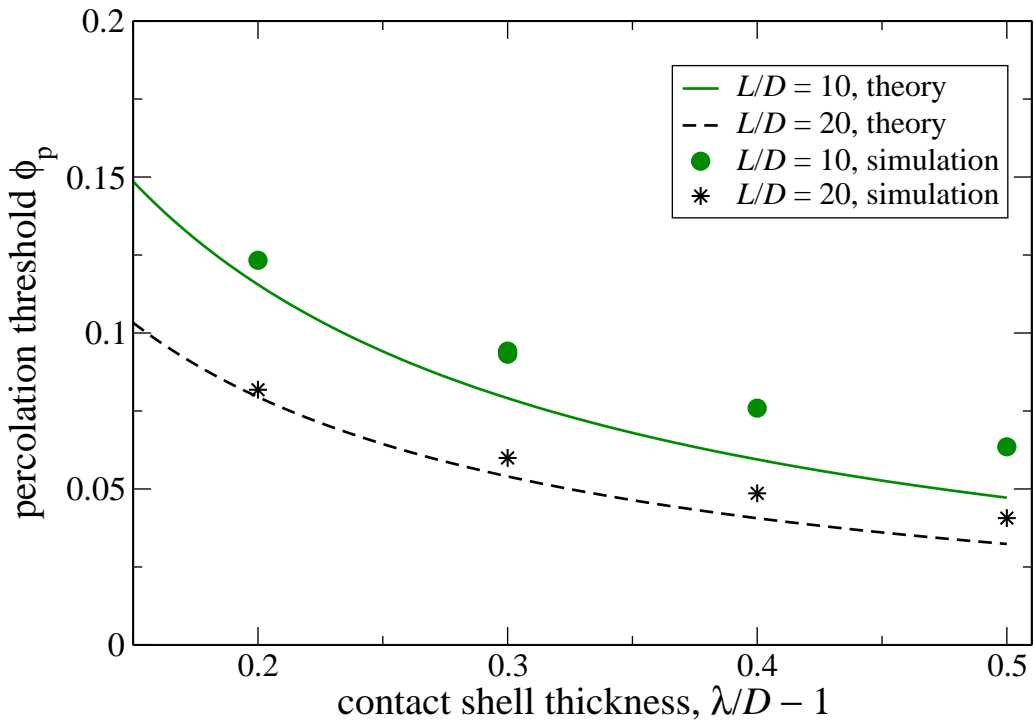}
  \caption{Percolation threshold as a function of the contact shell thickness,
i.e., the surface-to-surface distance criterion for connectivity.
    \label{fig:lambdaDependence}}
\end{figure}

Second, hard-core interactions seem to have a larger impact on the percolation threshold for large aspect ratios than for smaller ones. This observation agrees with previous Monte Carlo simulations of hard spheres \cite{bug:1985,Miller09c}. The smaller the connectivity range, the larger the difference between the percolation thresholds of hard and ideal rods. Fig.~\ref{fig:lambdaDependence} illustrates this point, showing a significant impact of the connectivity range on the percolation threshold even for relatively short rods of aspect ratio 10 and 20, itself quite accurately predicted by our theory.  In the slender rod limit, the ratio of percolation thresholds of ideal to hard rods should be proportional to the thickness of the connectivity shell around the core, as $(\lambda-D)/\lambda$.  The results in Fig.~\ref{fig:Allthresholds} bear out this theoretical prediction but, again, the limiting behavior is reached for the very longest of the rods included in our simulations.

\begin{figure}
  \centering
  \includegraphics[width=0.9\columnwidth]{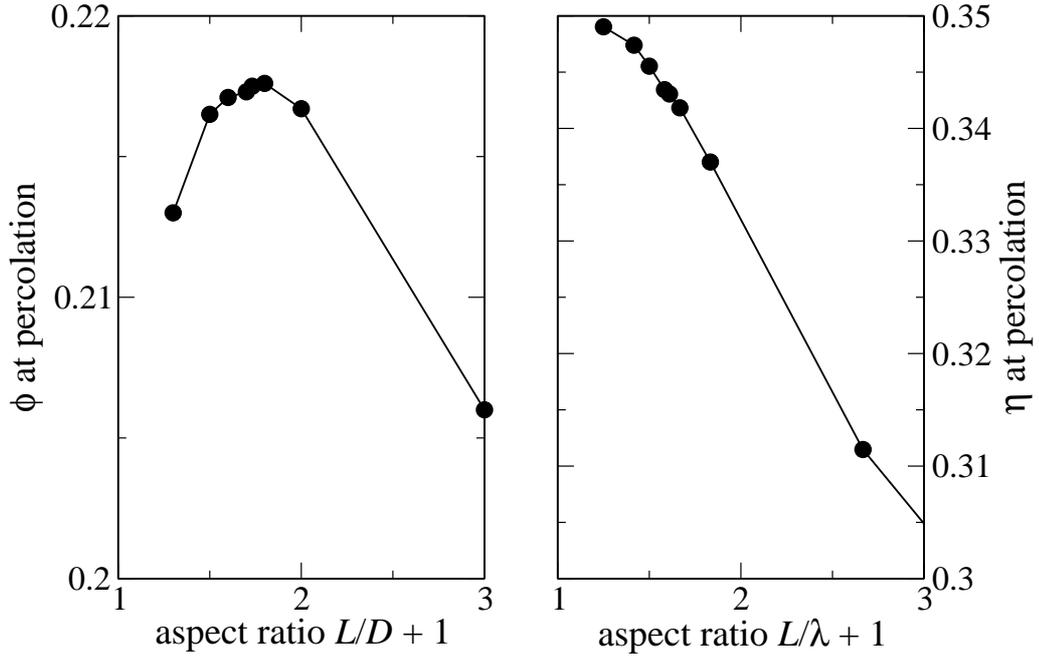}
  \caption{Percolation threshold for short hard rods with connectivity criterion $\lambda/D = 1.2$ from simulation.  If plotted as a function of the physical volume fraction of the hard cores (left-hand panel), the percolation threshold is non-monotonic in the aspect ratio, while the notional volume fraction of the connectivity shells (right-hand panel) shows no extremum.
  \label{fig:Shortrods}}
\end{figure}

In Fig.~\ref{fig:Shortrods} we focus on very short hard rods ($1.3 \le L/D+1 \le 3$). The connectivity criterion was again set to $\lambda/D = 1.2$. For short rods, if we define the percolation threshold in terms of the physical volume fraction $\phi_{\rm p}$ due to the cores of the particles, we find the threshold to be a non-monotonic function of the aspect ratio (left-hand panel of Fig.~\ref{fig:Shortrods}), with a maximum close to $L/D=0.7$ (full core aspect ratio $1.7$). Our theory also predicts a maximum at short aspect ratios. 
Maximisation of $\phi_{\rm p}$ in Eq.~(\ref{eq:eta}) with respect to $\gamma=L/D$ gives $L/D = \frac{2}{3}(-1+\sqrt{7+12\alpha+6\alpha^2})$ for the aspect ratio at maximum percolation threshold.  Using the connectivity range $\alpha=\lambda/D-1=0.2$ from the simulations, the theoretical expression evaluates to $L/D\approx1.4$, i.e., about a factor ot 2 larger than the simulation result.  One might be tempted to speculate that the maximum in $\phi_{\rm p}$ is related to the fact that slightly aspherical objects, such as ellipsoids, pack more efficiently than spheres \cite{donev:2004}.
However, if the data are plotted in terms of the renormalised volume fraction $\eta_{\rm p}$, the curve becomes monotonic (see Fig.~\ref{fig:Shortrods}, right-hand panel).
Thus the maximum arises from the definition of connectivity. Crucially, however, it is $\phi$ that corresponds to the experimental volume fraction of rod-like filler.  Hence, real composite materials should indeed show a local maximum in loading a the percolation threshold when short filler particles are used.  The renormalised volume fraction $\eta$ that includes the effective, penetrable contact shell is generally not accurately known {\it a priori}.

In summary, we have introduced a new version of connectedness percolation theory which accurately predicts percolation thresholds for nanorods over a large range of aspect ratios. We show that for aspect ratios below 1000 the percolation threshold deviates significantly from the inverse aspect ratio scaling prediction. Hence, caution is required when making predictions about systems in the intermediate regime of aspect ratios that is relevant to typical materials applications.
We have also presented new simulation data for hard and ideal rods, which are the first to cover aspect ratios from a sphere to very slender rods, providing a new benchmark. Ideal rods---a conveniently simple model---differ both quantitatively and qualitatively from more physical models that include an impenetrable core, and the model's limitations should therefore not be ignored in the context of real nanorod systems. Hard-core interactions change the approach to the theoretically expected scaling of the percolation threshold with the inverse aspect ratio and shift the percolation threshold to larger values than that for ideal rods by a factor that converges only slowly with respect to rod length. We have shown that the correction to the long-rod scaling regime that is needed at all realistic aspect ratios of hard nanorods can be obtained from connectedness percolation theory, demonstrating the strength of this approach.

\newpage
{\bf Supplemental Table 1:}
Percolation threshold of hard spherocylinders with cylindrical portion of length $L$, diameter $D$
and contact shell thickness $\lambda=1.2D$, from Monte Carlo simulations (blue circles in Figure 2 of the
article).
\begin{center}
\begin{tabular}{d{3.6}d{3.6}d{1.6}d{1.5}}
\toprule
\multicolumn{1}{c}{core aspect ratio} &
\multicolumn{1}{c}{overall aspect ratio} &
\multicolumn{1}{c}{percolation threshold} &
\multicolumn{1}{c}{uncertainty} \\
\multicolumn{1}{c}{$L/D$} &
\multicolumn{1}{c}{$L/\lambda+1$} &
\multicolumn{1}{c}{$\eta_{\rm p}$} &
\multicolumn{1}{c}{$\delta\eta_{\rm p}$} \\
\midrule
0.3 & 1.25 & 0.349 & 0.002 \\
0.5 & 1.416667 & 0.347 & 0.002 \\
0.6 & 1.5 & 0.346 & 0.002 \\
0.7 & 1.583333 & 0.343 & 0.002 \\
0.732051 & 1.610042 & 0.343 & 0.002 \\
0.8 & 1.666667 & 0.342 & 0.002 \\
1.0 & 1.833333 & 0.337 & 0.002 \\
2.0 & 2.666667 & 0.311 & 0.002 \\
10.0 & 9.333333 & 0.17974 & 0.00009 \\
20.0 & 17.666667 & 0.11848 & 0.00003 \\
30.0 & 26 & 0.08892 & 0.00002 \\
50.0 & 42.666667 & 0.05930 & 0.00007 \\
70.0 & 59.333333 & 0.0447 & 0.0003 \\
100.0 & 84.333333 & 0.03250 & 0.00004 \\
200.0 & 167.666667 & 0.0171 & 0.0004 \\
1000.0 & 834.333333 & 0.00359 & 0.00006 \\
\bottomrule
\end{tabular}
\end{center}

\newpage
{\bf Supplemental Table 2:}
Percolation threshold of ideal spherocylinders with cylindrical portion of length $L$ and diameter $\lambda$
from Monte Carlo simulations (red pluses in Figure 2 of the article).  The fractional uncertainty in
the percolation threshold is $\delta\eta_{\rm p}/\eta_{\rm p}=0.001$
\begin{center}
\begin{tabular}{d{4.1}d{1.7}}
\toprule
\multicolumn{1}{c}{overall aspect ratio} &
\multicolumn{1}{c}{percolation threshold} \\
\multicolumn{1}{c}{$L/\lambda+1$} &
\multicolumn{1}{c}{$\eta_{\rm p}$} \\
\midrule
1 & 0.3424 \\
1.1 & 0.3403 \\
1.2 & 0.3373 \\
1.3 & 0.3320 \\
1.4 & 0.3256 \\
1.5 & 0.3188 \\
1.6 & 0.3122 \\
1.8 & 0.2969 \\
2 & 0.2823 \\
2.2 & 0.2687 \\
2.5 & 0.2495 \\
3 & 0.2215 \\
4 & 0.1772 \\
5 & 0.1464 \\
7 & 0.1072 \\
9 & 0.08368 \\
11 & 0.06814 \\
16 & 0.04564 \\
21 & 0.03389 \\
31 & 0.02208 \\
41 & 0.01619 \\
51 & 0.01279 \\
61 & 0.01044 \\
81 & 0.007636 \\
101 & 0.005992 \\
151 & 0.003878 \\
201 & 0.002846 \\
401 & 0.001366 \\
601 & 0.0008955 \\
1001 & 0.0005281 \\
2001 & 0.0002590 \\
\bottomrule
\end{tabular}
\end{center}

%

\end{document}